\documentclass[12pt]{article}

\usepackage{graphicx}
\usepackage{dcolumn}
\usepackage{bm}
\usepackage{cite}
\usepackage{amsmath}
\usepackage{amssymb}

\begin{document}

\begin{center}
\Large\bf\boldmath
\vspace*{0.8cm} SUSY Constraints, Relic Density,\\ and Very Early Universe
\unboldmath
\end{center}

\vspace{0.6cm}
\begin{center}
A. Arbey$^{1,}$\footnote{Electronic address: \tt alexandre.arbey@ens-lyon.fr} and F. Mahmoudi$^{2,}$\footnote{Electronic address: \tt mahmoudi@in2p3.fr}\\[0.4cm]
{\sl $^1$ Universit\'e de Lyon, Lyon, F-69000, France; Universit\'e Lyon~1, Villeurbanne, \\
F-69622, France; Centre de Recherche Astrophysique de Lyon, Observatoire de Lyon, \\
9 avenue Charles Andr\'e, Saint-Genis Laval cedex, F-69561, France; CNRS, UMR 5574; \\
Ecole Normale Sup\'erieure de Lyon, Lyon, France.}\\
\vspace{0.6cm}
{\sl $^2$ Clermont Universit\'e, Universit\'e Blaise Pascal, CNRS/IN2P3,\\ LPC, BP 10448, 63000 Clermont-Ferrand, France.}
\end{center}

\vspace{0.4cm}
\begin{abstract}
\noindent 
The sensitivity of the lightest supersymmetric particle relic density calculation to different cosmological scenarios is discussed. In particular, we investigate the effects of modifications of the expansion rate and of the entropy content in the Early Universe. These effects, even with no observational consequences, can still drastically modify the relic density constraints on the SUSY parameter space. We suggest general parametrizations to evaluate such effects, and derive also constraints from Big-Bang nucleosynthesis. We show that using the relic density in the context of supersymmetric constraints requires a clear statement of the underlying cosmological model assumptions to avoid misinterpretations. On the other hand, we note that combining the relic density calculation with the eventual future discoveries at the LHC will hopefully shed light on the Very Early Universe properties.
\end{abstract}


\section{Introduction}

Cosmological observations reveal that the mass content of the Universe is mostly composed of dark matter of unknown, but non-baryonic nature. New physics models, such as supersymmetry (SUSY), provide stable particle candidates for dark matter, and one can compute their present energy density, the relic density \cite{relic_calculation}. This relic density is often compared to the dark matter density deduced from cosmological observations in order to constrain new physics parameters (see for example \cite{relic_constraints}). The usual assumption in doing that is that the Universe is ruled by the standard model of cosmology, which assumes that radiation energy density and radiation entropy density dominate the Universe properties in the Very Early Universe. However, before Big-Bang nucleosynthesis (BBN) many phenomena could have modified the physical properties of the Universe, such as the expansion rate, or the entropy evolution, or even the non-thermal production of relic particles. The calculation of the relic density is altered in these cases, and for example the influence of a quintessence-like scalar field \cite{quintessence}, or reheating and non thermal production of relic particles due to the decay of an inflaton-like scalar field \cite{reheating} have been discussed in the literature. Scenarios involving dark fluids or extra-dimensions modifying the expansion rate of the Universe have been also considered in \cite{arbey}. On the other hand, with the start-up of the LHC, we can hope for new information on the physics beyond the standard model, providing more hints for the determination of the nature of dark matter \cite{inverse}.

In this paper, we consider the calculation of the relic density beyond the cosmological standard model and propose a generalized parametrization of the modification of the entropy evolution in the Early Universe. Effects of the modification of the expansion rate was studied in \cite{arbey}, and we derive here the necessary equations to compute the relic density in a more generic way. We also discuss the importance of the LHC data in the context of cosmology and claim that future discoveries of the LHC can lead to a better understanding of the Universe properties before BBN. 

To illustrate the consequences of modifications of the cosmological model on the calculation of the relic density we consider in the following the minimal supersymmetric extension of the Standard Model (MSSM) with $R$-parity conservation and show the implications on the SUSY parameter interpretation and constraints.

\section{Relic density calculation}

The density number of supersymmetric particles is determined by the Boltzmann equation, which in presence of non-thermal production of relic particles takes the form:
\begin{equation}
\frac{dn}{dt} = - 3 H n - \langle \sigma v \rangle (n^2 - n^2_{eq}) + N_D  \;,\label{boltzmann}
\end{equation}
where $n$ is the number density of supersymmetric particles, $\langle \sigma v \rangle$ is the thermally averaged annihilation cross-section, $H$ is the Hubble expansion rate and $n_{eq}$ is the supersymmetric particle equilibrium number density. The term $N_D$ provides a parametrization of the non-thermal production of SUSY particles which is in general temperature-dependent. The expansion rate $H$ is determined by the Friedmann equation:
\begin{equation}
 H^2=\frac{8 \pi G}{3} (\rho_{rad} + \rho_D)  \;.\label{friedmann}
\end{equation}
$\rho_{rad}$ is the radiation energy density, which is considered to be dominant before BBN in the standard cosmological model. Following \cite{arbey}, we introduced in Eq.~(\ref{friedmann}) $\rho_D$ to parametrize the expansion rate modification. $\rho_D$ can be interpreted either as an additional energy density term modifying the expansion ({\it e.g.} quintessence), or as an effective energy density which can account for other phenomena affecting the expansion rate ({\it e.g.} extra-dimensions).

The entropy evolution can also be altered beyond the standard cosmological model, and we write the entropy evolution equation in presence of entropy fluctuations:
\begin{equation}
\frac{ds}{dt} = - 3 H s + \Sigma_D \label{entropy_evolution} \;,
\end{equation}
where $s$ is the total entropy density. $\Sigma_D$ in the above equation parametrizes effective entropy fluctuations due to unknown properties of the Early Universe, and is temperature-dependent.

In the following, for the sake of generality, the three new parameters $N_D$, $\rho_D$ and $\Sigma_D$ are regarded as independent. Entropy and energy alterations are considered here as effective effects, which can be generated by curvature, phase transitions, extra-dimensions, or other phenomena in the Early Universe. In a specific physical scenario, these parameters may be related, as for example in reheating models \cite{reheating}. However, the large number of unanswered questions in the pre-BBN epoch and the complexity of particle physics models, which involves many different fields, can doubt the simplicity of reheating models. In particular many open questions remain in inflation, leptogenesis and baryogenesis scenarios. Therefore, a complete and realistic description of the Early Universe would rely on several different fields with complementary roles, far beyond the decaying scalar field description of inflation and reheating models, and the direct dependence between energy and entropy would in such cases be very difficult to determine at the time of the relic freeze-out. For this reason, we prefer to adopt a more conservative and effective approach in which the effective energy and entropy densities are considered as independent.

The radiation energy and entropy densities can be written as usual:
\begin{equation}
\rho_{rad}=g_{eff}(T) \frac{\pi^2}{30} T^4 \;, \qquad\qquad s_{rad} = h_{eff}(T) \frac{2\pi^2}{45} T^3 \;. \label{srad}
\end{equation}
We split the total entropy density into two parts: radiation entropy density and effective dark entropy density, $s \equiv s_{rad} + s_D$. Using Eq.~(\ref{entropy_evolution}) the relation between $s_D$ and $\Sigma_D$ can then be derived:
\begin{equation}
\Sigma_D = \sqrt{\frac{4 \pi^3 G}{5}} \sqrt{1 + \tilde{\rho}_D} T^2 \left[\sqrt{g_{eff}} s_D - \frac13  \frac{h_{eff}}{g_*^{1/2}} T \frac{ds_D}{dT}\right] \;, \label{SigmaD}
\end{equation}
with
\begin{equation}
g_*^{1/2} = \frac{h_{eff}}{\sqrt{g_{eff}}}\left(1+\frac{T}{3 h_{eff}} \frac{dh_{eff}}{dT}\right) \;.
\end{equation}
Following the standard relic density calculation method \cite{relic_calculation}, $Y \equiv n/s$ is introduced, and Eq.~(\ref{boltzmann}) yields
\begin{equation}
 \frac{dY}{dx}= - \frac{m_{lsp}}{x^2} \sqrt{\frac{\pi}{45 G}} g_*^{1/2} \left( \frac{1 + \tilde{s}_D}{\sqrt{1+\tilde{\rho}_D}} \right) \left[\langle \sigma v \rangle (Y^2 - Y^2_{eq}) + \frac{Y \Sigma_D - N_D}{\left(h_{eff}(T) \frac{2\pi^2}{45} T^3\right)^2 (1+\tilde{s}_D)^2} \right] \;, \label{final}
\end{equation}
where $x=m_{lsp}/T$, $m_{lsp}$ is the mass of the lightest supersymmetric relic particle, and
\begin{equation}
 \tilde{s}_D \equiv \frac{s_D}{h_{eff}(T) \frac{2\pi^2}{45} T^3}\;, \qquad\qquad \tilde{\rho}_D \equiv \frac{\rho_D}{g_{eff} \frac{\pi^2}{30} T^4}\;,
\end{equation}
and
\begin{equation}
 Y_{eq} = \frac{45}{4 \pi^4 T^2} h_{eff} \frac{1}{(1+\tilde{s}_D)} \sum_i g_i m_i^2 K_2\left(\frac{m_i}{T}\right) \;,
\end{equation}
with $i$ running over all supersymmetric particles of mass $m_i$ and with $g_i$ degrees of freedom. $s_D$ and $\Sigma_D$ are not independent variables and are related through Eq.~(\ref{SigmaD}). In the limit where $\rho_D = s_D = \Sigma_D = N_D = 0$, standard relations are retrieved. Integrating Eq. (\ref{final}), the relic density can then be calculated using:
\begin{equation}
 \Omega h^2 = \frac{m_{lsp} s_0 Y_0 h^2}{\rho_c^0} = 2.755 \times 10^8 Y_0 m_{lsp}/\mbox{GeV} \;, \label{omegah2}
\end{equation}
where the subscript 0 refers to the present values of the parameters. Using Eqs. (\ref{boltzmann}-\ref{omegah2}) the relic density in presence of a modified expansion rate, of entropy fluctuations and of non-thermal production of relic particles, can be computed provided $\rho_D$, $N_D$ and $s_D$ or $\Sigma_D$ are specified.
For $\rho_D$ we follow the parametrization introduced in Ref. \cite{arbey}:
\begin{equation}
 \rho_D =  \kappa_\rho \rho_{rad}(T_{BBN}) \bigl(T/T_{BBN}\bigr)^{n_\rho} \;, \label{rhoD}
\end{equation}
where $T_{BBN}$ is the BBN temperature. Different values of $n_\rho$ leads to different behaviors of the effective density. For example, $n_\rho=4$ corresponds to a radiation behavior, $n_\rho=6$ to a quintessence behavior, and $n_\rho>6$ to the behavior of a decaying scalar field. $\kappa_\rho$ is the ratio of the effective energy density to the radiation energy density at BBN time and can be negative. The role of $\rho_D$ is to increase the expansion rate for $\rho_D > 0$, leading to an early decoupling and a higher relic density, or to decrease it for $\rho_D < 0$, leading to a late decoupling and to a smaller relic density. Requiring that the radiation density dominates during BBN implies $|\kappa_\rho| \ll 1$. Moreover, $H^2 > 0$ imposes $|\rho_D| < \rho_{rad}$ for $\rho_D < 0$, strongly limiting the interest of negative $\kappa_\rho$ as mentioned in \cite{arbey}.

To model the entropy perturbations, we choose to parametrize $s_D$ in a similar way\footnote{An alternative parametrization, more similar to the one used in reheating models \cite{reheating}, can be done by choosing $\Sigma_D$ instead of $s_D$.}:
\begin{equation}
 s_D =  \kappa_s s_{rad}(T_{BBN}) \bigl(T/T_{BBN}\bigr)^{n_s} \;. \label{sD}
\end{equation}
This parametrization finds its roots in the first law of thermodynamics, where energy and entropy are directly related and therefore the entropy parametrization can be similar to the energy parametrization. As for the energy density, different values of $n_s$ lead to different behaviors of the entropy density: $n_s = 3$ corresponds to a radiation behavior, $n_s = 4$ appears in dark energy models, $n_s \sim 1$ in reheating models, and other values can be generated by curvature, scalar fields or extra-dimension effects. $\kappa_s$ is the ratio of the effective entropy density to the radiation entropy density at BBN time and can be negative. The role of $s_D$ will be to increase the temperature at which the radiation dominates for $s_D > 0$, leading to a decreased relic density, or to decrease this temperature for $s_D < 0$, leading to an increased relic density. For naturalness reason, we impose that the radiation entropy density dominates at BBN time, {\it i.e.} $|\kappa_s| \ll 1$. Constraints on the cosmological entropy in reheating models are derived in \cite{entropy_constraints}, and we extend here the analyses to the general parametrization (\ref{sD}) using BBN data. 

A general parametrization is difficult for $N_D$. In many reheating models a scalar field decays into supersymmetric particles, and the non-thermal production is therefore related to the scalar field density. To avoid imposing {\it ad hoc} general conditions, we choose $N_D=0$. We can however note that the main effect of the non-thermal production is an enhancement of the final number of relic particles, so that it is always possible to enhance the final relic density by assuming non-thermal production of relic particles.

\section{BBN constraints}
\begin{figure}[!t]
\begin{center}
\includegraphics[width=8cm]{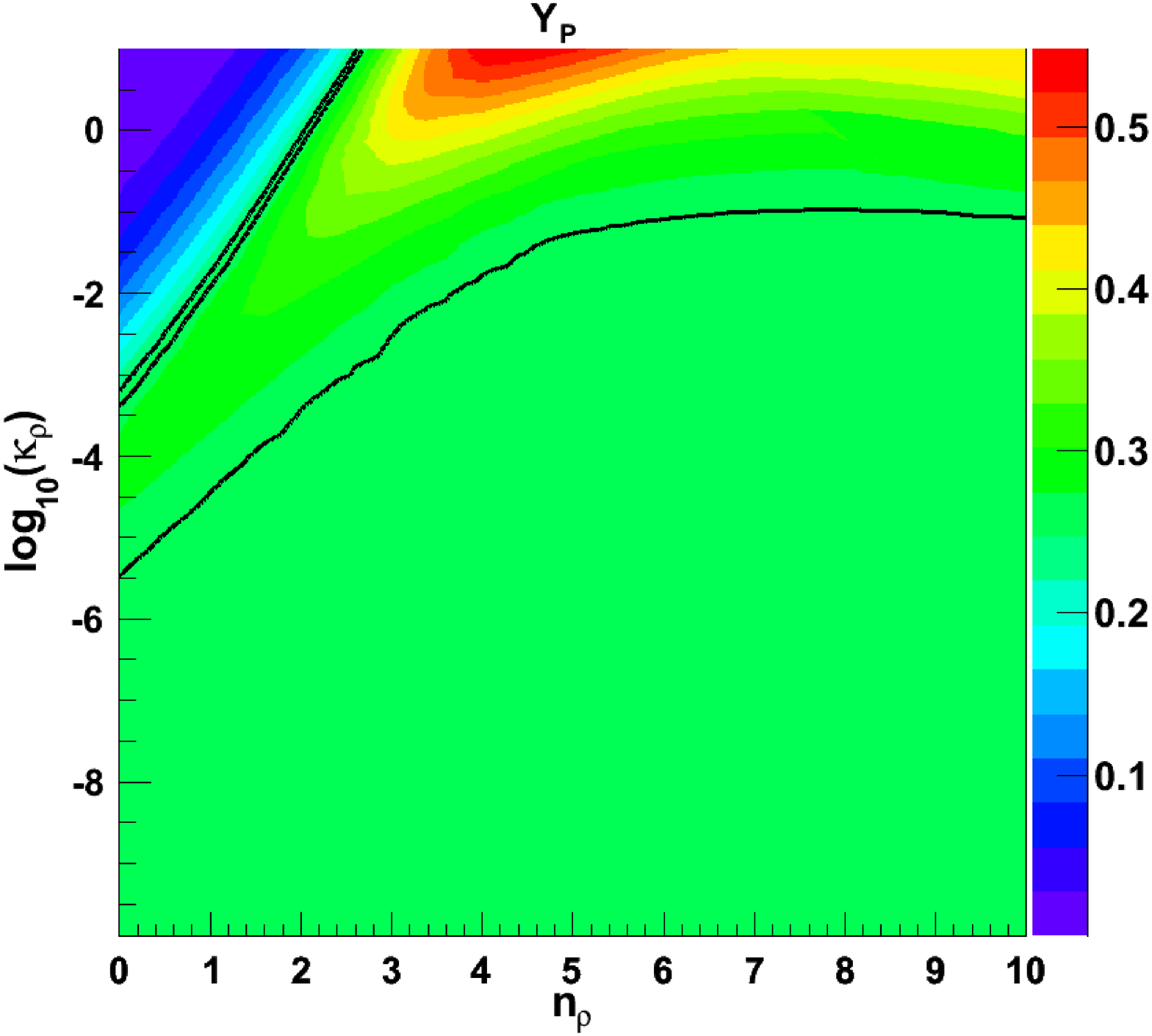}\hspace*{0.2cm}\includegraphics[width=8cm]{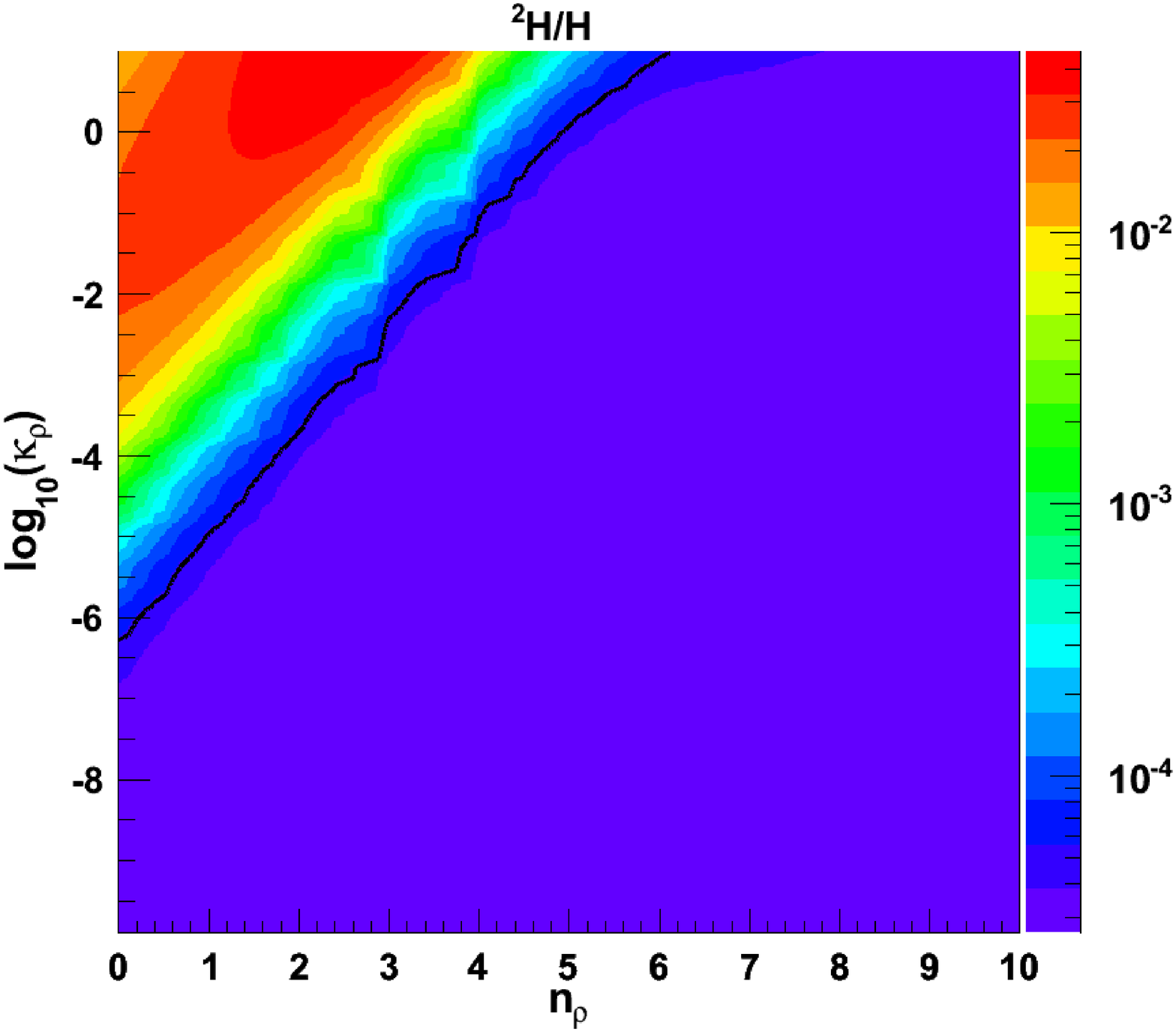}
\end{center}
\caption{Constraints from $Y_p$ (left) and $^2\!H/H$ (right) on the effective dark energy. The parameter regions excluded by BBN are located above the black lines. The colors correspond to different values of $Y_p$ and $^2\!H/H$.\label{bbnfig}}
\end{figure}
\begin{figure}[!t]
\begin{center}
\includegraphics[width=8cm]{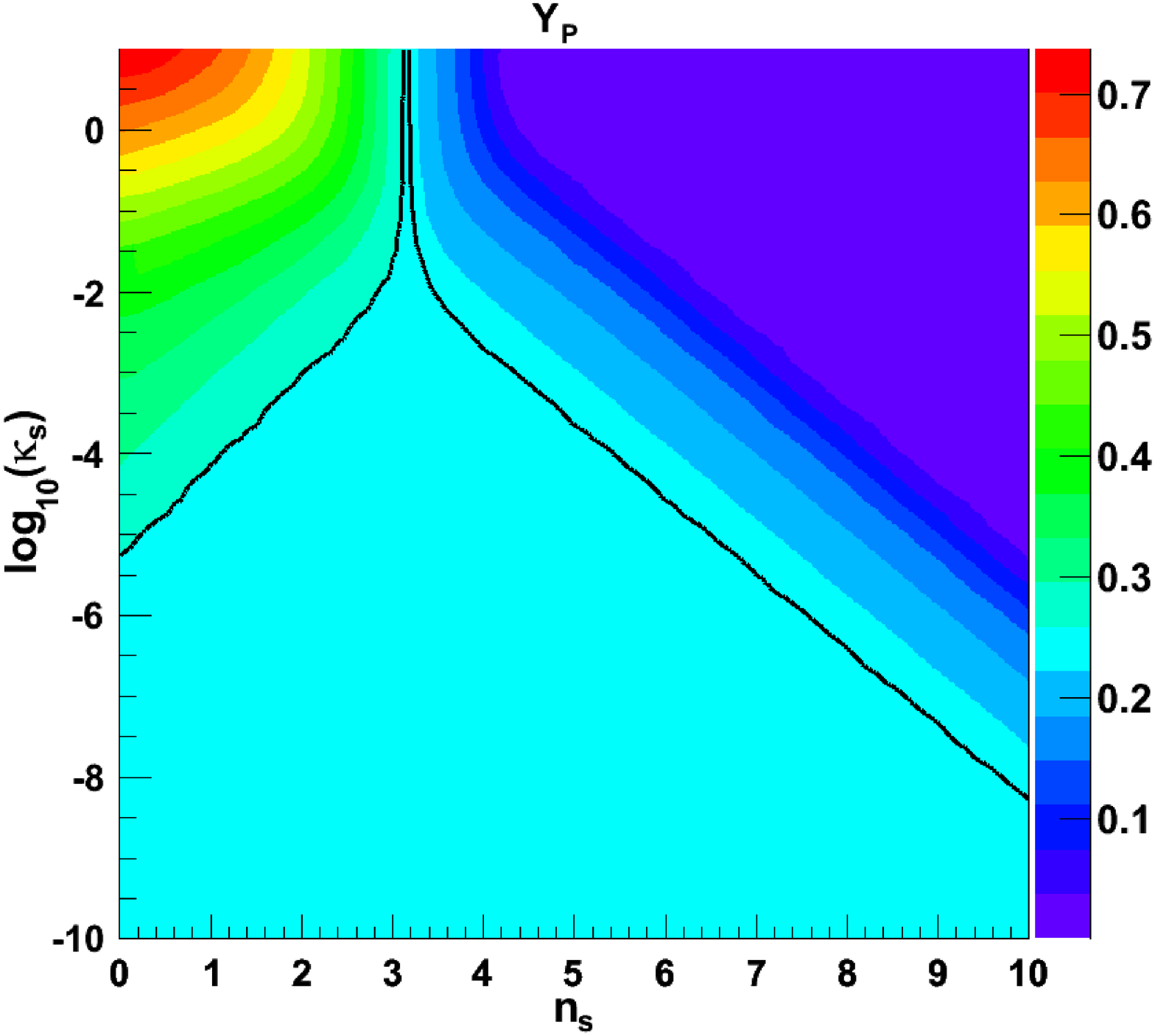}\hspace*{0.2cm}\includegraphics[width=8cm]{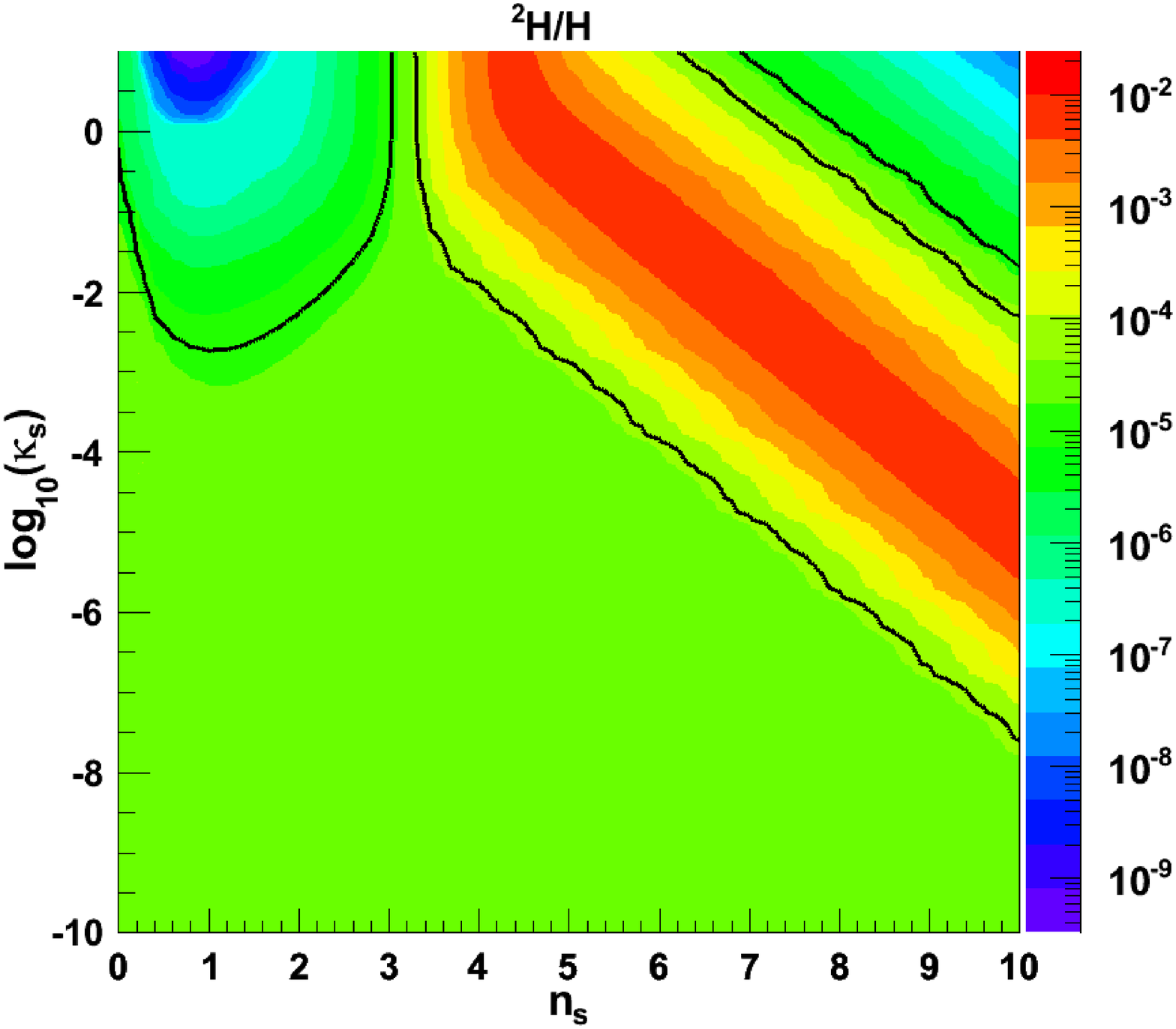}
\end{center}
\caption{Constraints from $Y_p$ (left) and $^2\!H/H$ (right) on the effective dark entropy. The parameter regions excluded by BBN are located above the black lines. The colors correspond to different values of $Y_p$ and $^2\!H/H$.\label{bbnfig2}}
\end{figure}

%
In order to make a realistic analysis of the allowed cosmological scenarios, we apply the BBN constraints. To compute the relevant abundances of the elements, we use a version of the BBN abundance calculation code NUC123 \cite{kawano92} updated with the NACRE \cite{nacre} reaction compilation and modified to include the parametrization of the expansion rate and entropy content of Eqs. (\ref{friedmann}), (\ref{rhoD}) and (\ref{sD}). We consider the rather conservative bounds of \cite{jedamzik06}:
\begin{eqnarray}
0.240 < Y_p < 0.258\;, \qquad 1.2 \times 10^{-5} < \!~^2\!H/H < 5.3 \times 10^{-5}\;,\qquad\qquad\\
\nonumber 0.57 < \!~^3\!H/\!~^2\!H < 1.52\;, \qquad  \!~^7\!Li/H > 0.85 \times 10^{-10}\;,\qquad \!~^6\!Li/\!~^7\!Li < 0.66\;,
\end{eqnarray}
for the helium abundance $Y_p$ and the primordial $^2\!H/H$, $^3\!H/\!~^2\!H$, $^7\!Li/H$ and $^6\!Li/\!~^7\!Li$ ratios. The most constraining observables are $Y_p$ and $^2\!H/H$, and the constraints obtained are shown in Fig.~\ref{bbnfig} for ($\kappa_\rho$, $n_\rho$), and in Fig.~\ref{bbnfig2} for ($\kappa_s$, $n_s$).%

The BBN constraints can be summarized as:
\begin{equation}
\kappa_\rho \lesssim 10^{-1.5}\;, \qquad\qquad \kappa_\rho \lesssim 10^{1.2 n_\rho - 6.0} \;,
\end{equation}
\begin{equation}
\kappa_s \lesssim 10^{n_s -5.2}\;, \qquad\qquad \kappa_s \lesssim 10^{-0.8 n_s + 0.5} \;.
\end{equation}
Also, for consistency with the CMB observations, we impose either $n_\rho \ge 4$ and $n_s \ge 3$, or $\rho_D = s_D = 0$ for $T < T_{BBN}$, so that $\rho_D$ and $s_D$ do not dominate after BBN.

\section{SUSY constraints}
We now consider the influence of the modified cosmological model on the supersymmetric constraints. The following computations are performed with SuperIso Relic v2.7 \cite{superiso,superiso_relic}. Considering the latest WMAP data \cite{WMAP5} with an additional 10\% theoretical uncertainty on the relic density calculation, we derive the following favored interval at 95\% C.L.:
\begin{equation}
 0.088 < \Omega_{DM} h^2 < 0.123 \;.
\end{equation}
The older dark matter interval is also considered:
\begin{equation}
 0.1 < \Omega_{DM} h^2 < 0.3 \;.\label{old}
\end{equation}
In the following, we restrict the parameters to $n_\rho \ge 4$, $0 \le \kappa_\rho \le 1$, $n_s \ge 3$, $0 \le \kappa_s \le 1$, and consider a constrained MSSM scenario. To allow more flexibility in the Higgs sector, we focus on the Non-Universal Higgs Mass Model (NUHM), in which the parameters consist of the universal (non-Higgs) scalar mass at GUT scale $m_0$, the universal gaugino mass at GUT scale $m_{1/2}$, the trilinear soft breaking parameter at GUT scale $A_0$, the ratio of the two Higgs vacuum expectation values $\tan\beta$, the $\mu$ parameter and the CP-odd Higgs mass $m_A$.

We consider first the NUHM test-point ($m_0=m_{1/2}=1$ TeV, $m_A=\mu=500$ GeV, $A_0=0$, $\tan\beta=40$) which gives a relic density of $\Omega h^2 \approx 0.11$, favored by the WMAP constraints.

\begin{figure}[!t]
\begin{center}
\hspace*{-0.5cm}\includegraphics[width=6.cm]{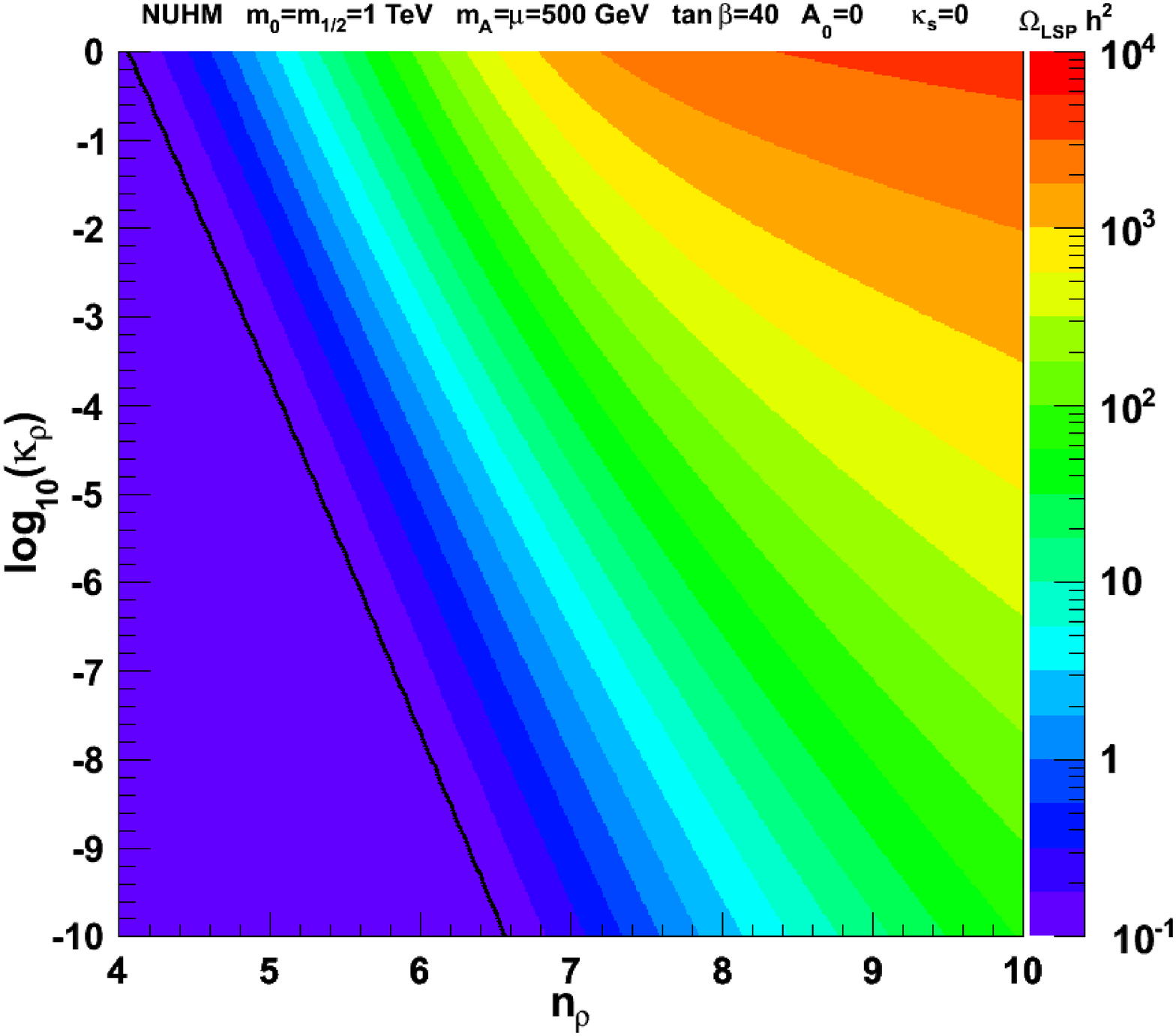}\includegraphics[width=6.cm]{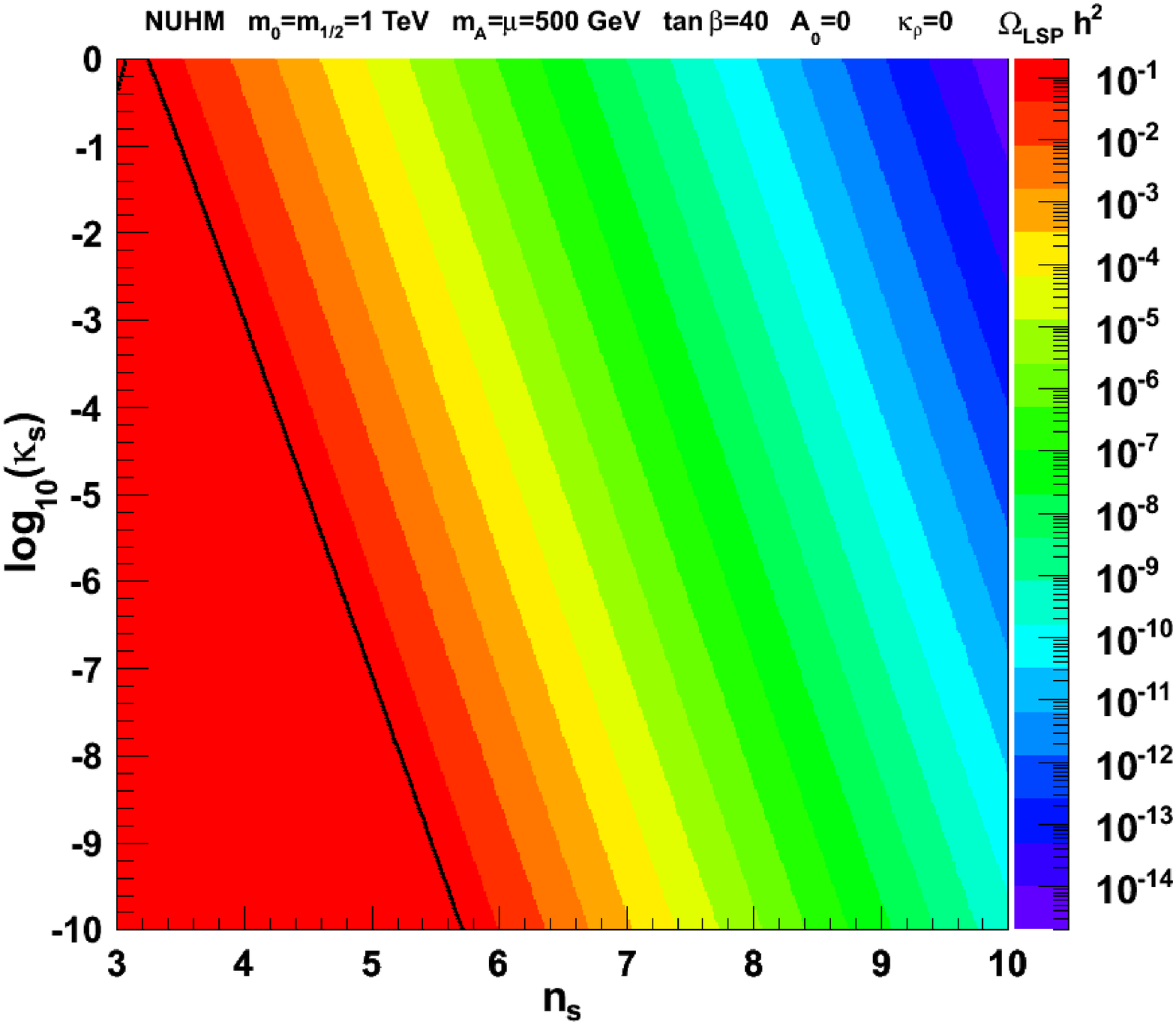}\includegraphics[width=6.cm]{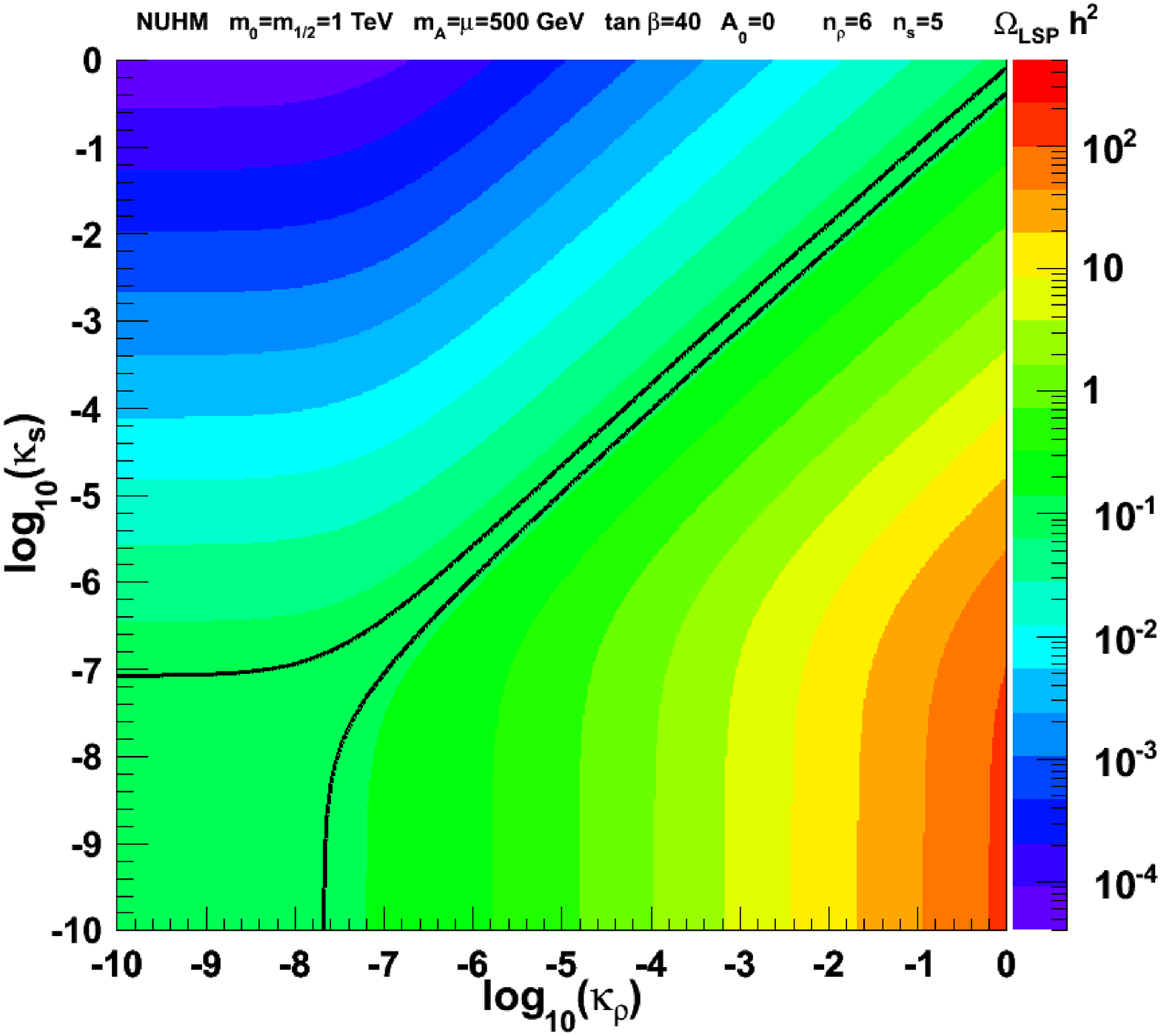}
\end{center}
\caption{Influence of the presence of an effective energy density (left), an effective entropy (center), and both an effective energy with $n_\rho=6$ and an effective entropy with $n_s=5$ (right). The colors correspond to different values of $\Omega h^2$. The black lines delimit the regions favored by WMAP. The favored zones are the lower left corners for the first two plots, and between the black lines for the last plot.\label{fig3}}
\end{figure}

Three different effects are presented in Fig. \ref{fig3}: the first plot shows the influence of the presence of an additional effective density on the computed relic density. We note that when $\kappa_\rho$ and $n_\rho$ increase, the relic density increases up to a factor of $10^5$, as already noticed in \cite{arbey}. The second plot illustrates the effect of an additional entropy density, in absence of additional energy density. Here when $\kappa_s$ and $n_s$ increase, the relic density is strongly decreased down to a factor of $10^{-14}$. The third plot is an example of both additional energy density with $n_\rho=6$ and additional entropy density with $n_s=5$. In this case, the values of the relic density varies from $10^{-4}$ to $10^3$, and we notice a narrow strip between the WMAP lines in which the entropy and energy effects almost cancel, leading to a degenerate zone with $\Omega h^2 \approx 0.11$.
\begin{figure}[!h]
\begin{center}
\includegraphics[width=6.45cm]{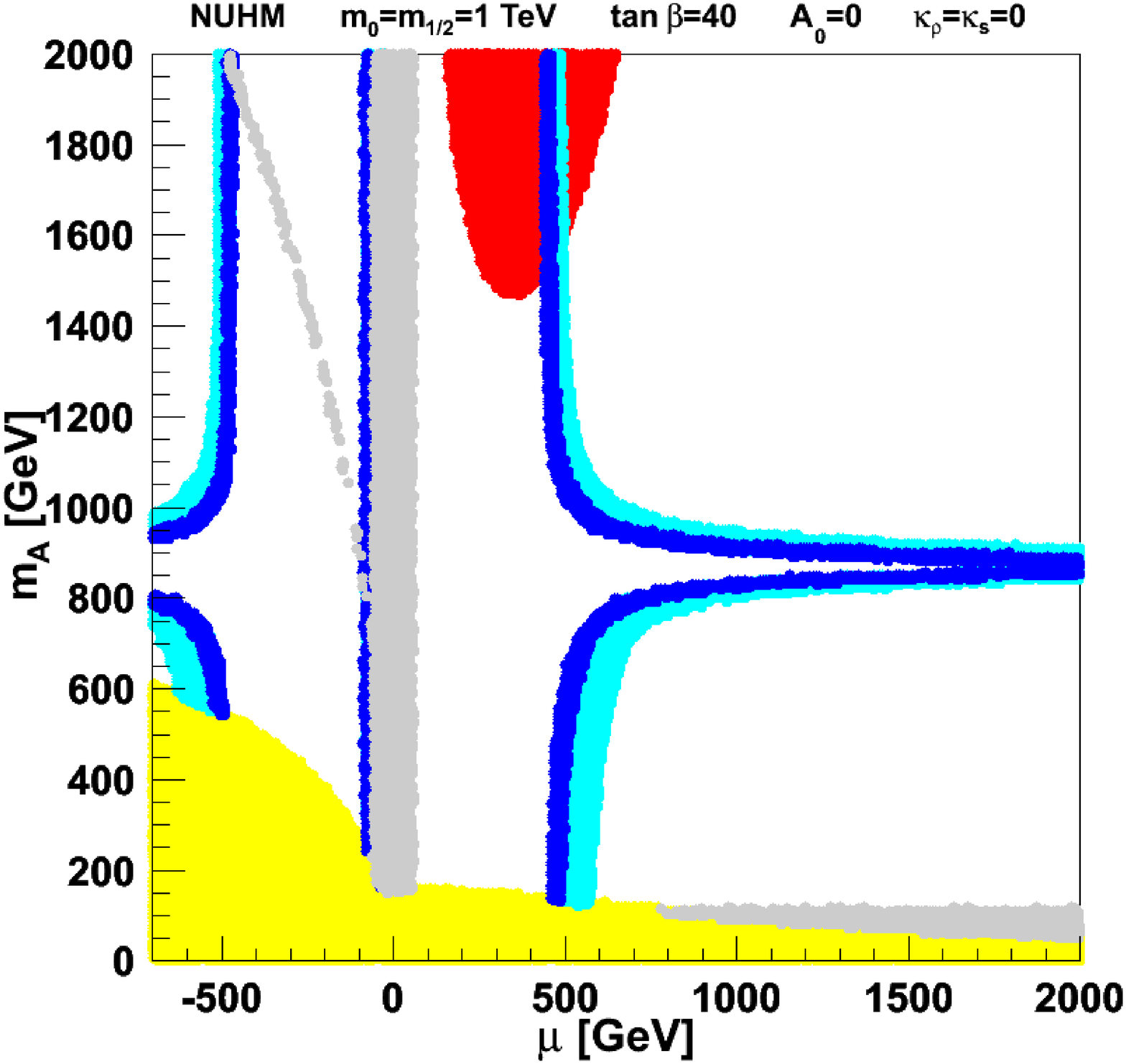}\hspace*{0.2cm}\includegraphics[width=6.45cm]{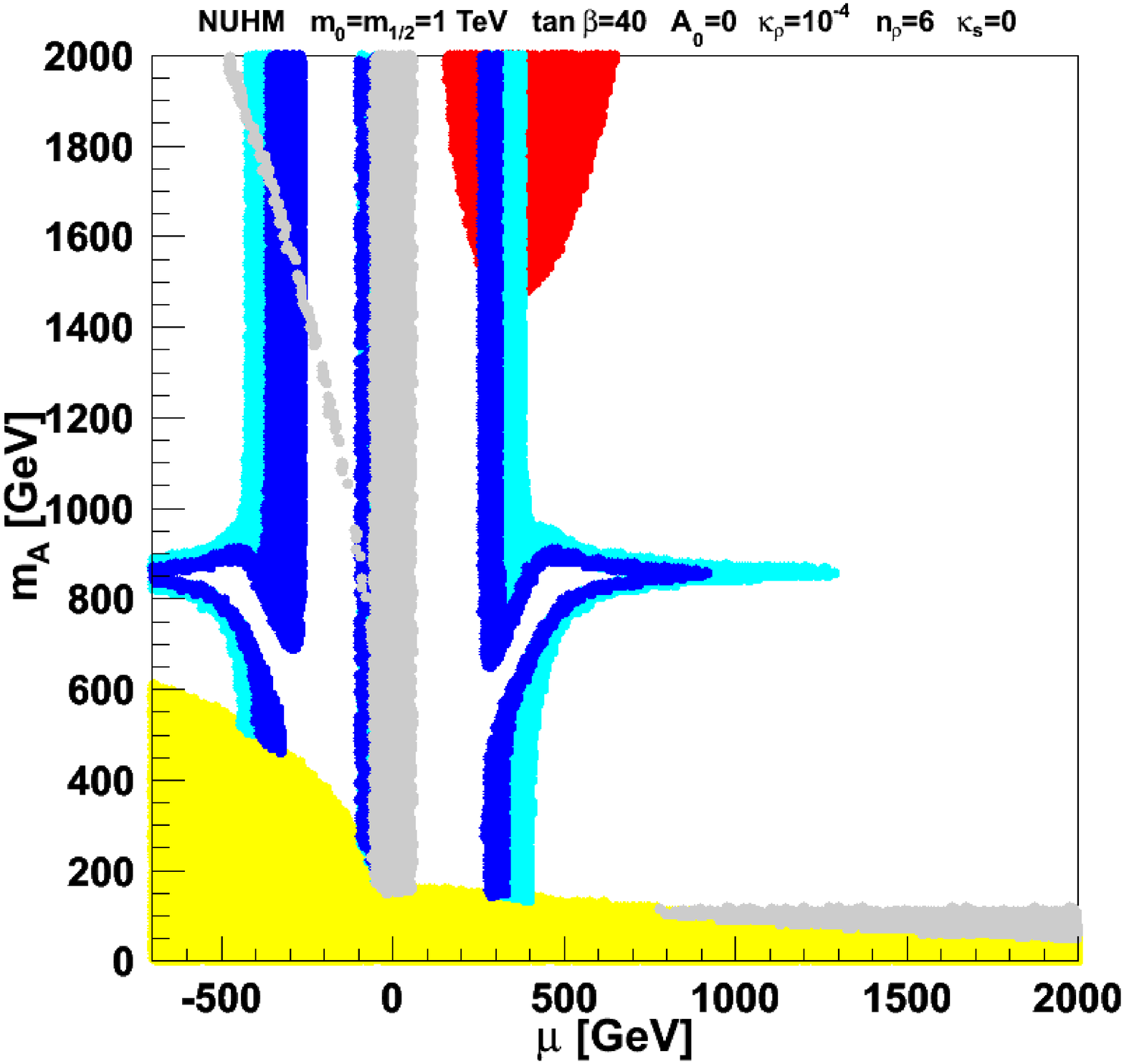}
\includegraphics[width=6.45cm]{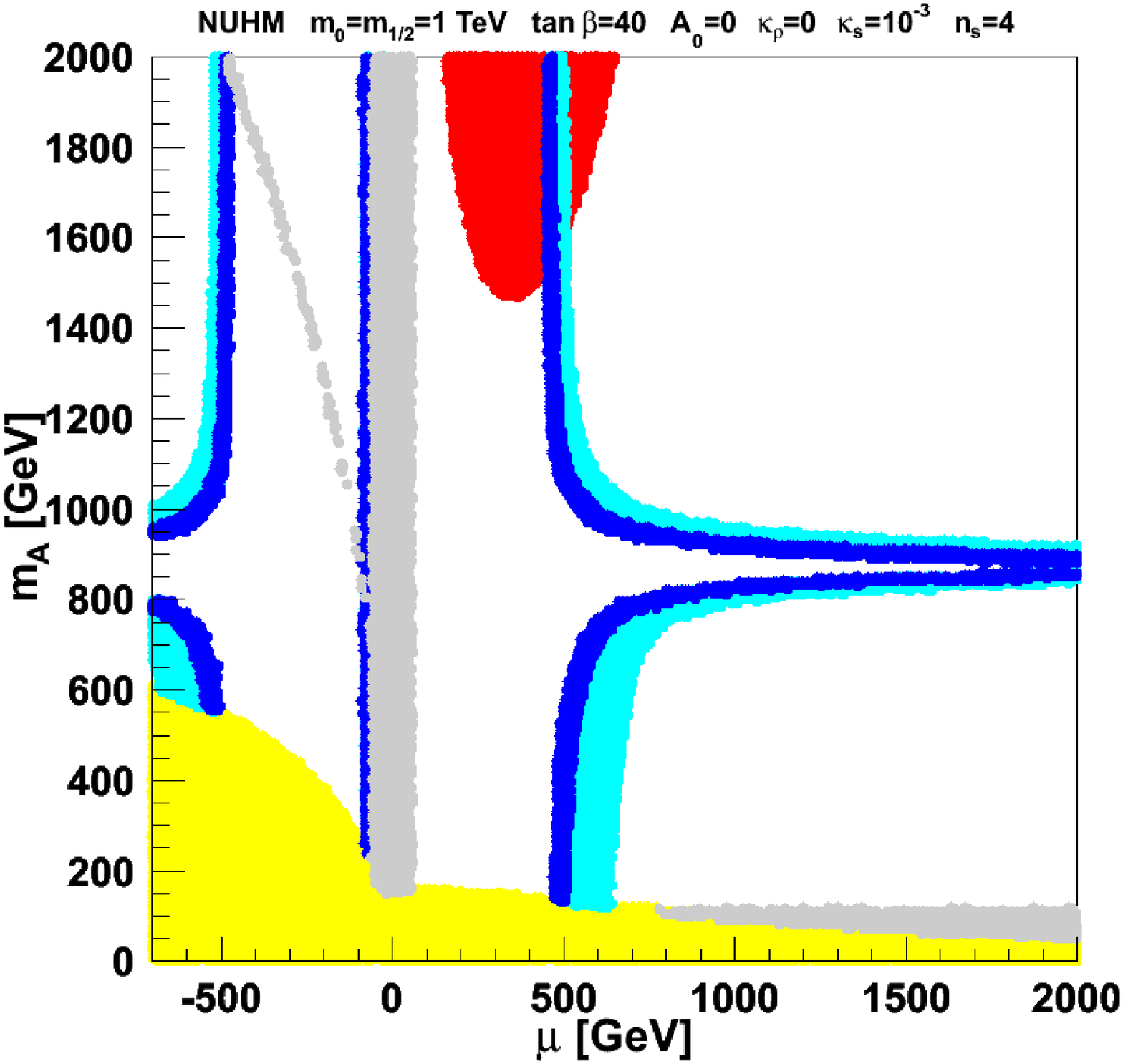}\hspace*{0.2cm}\includegraphics[width=6.45cm]{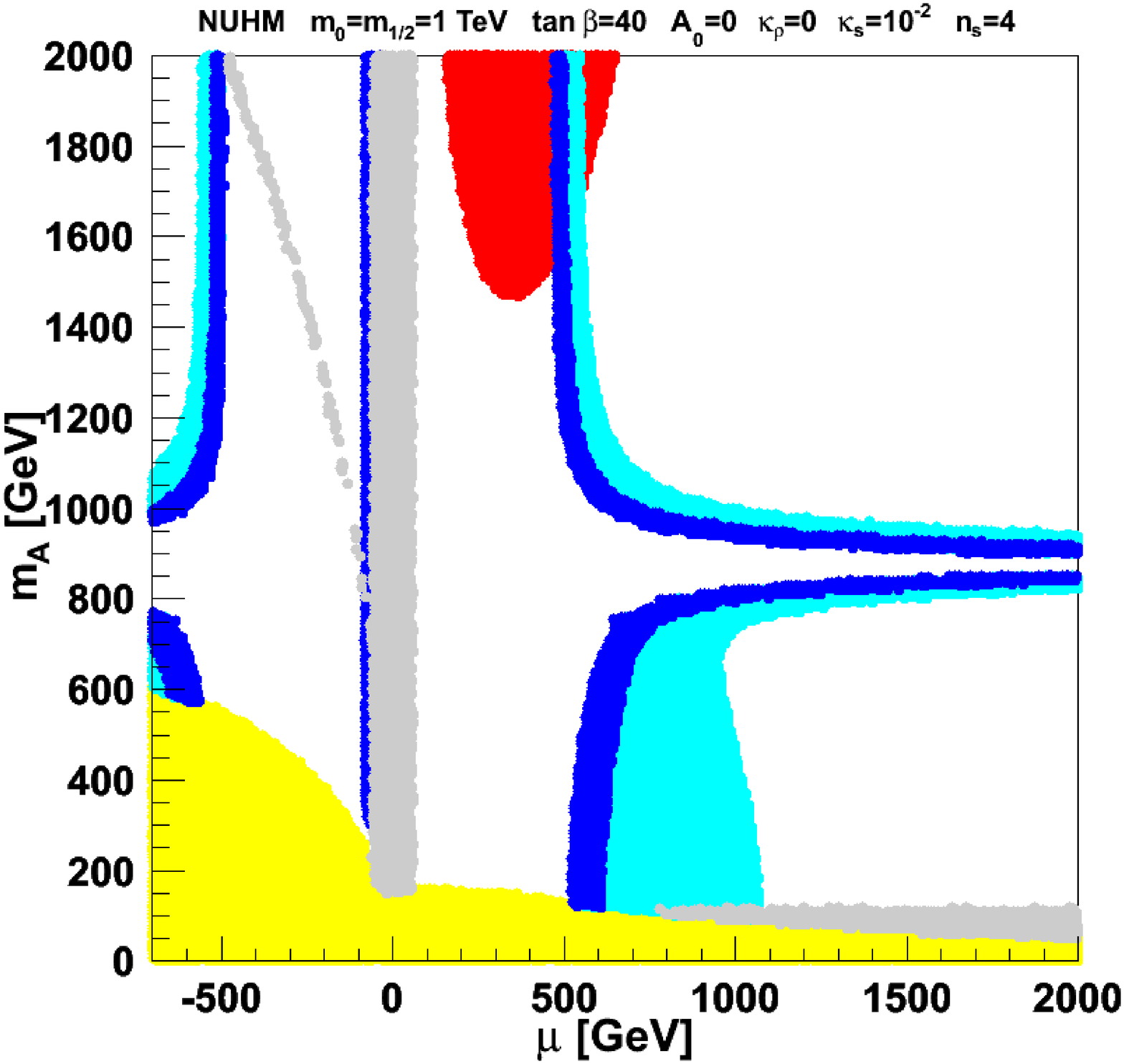}
\includegraphics[width=6.45cm]{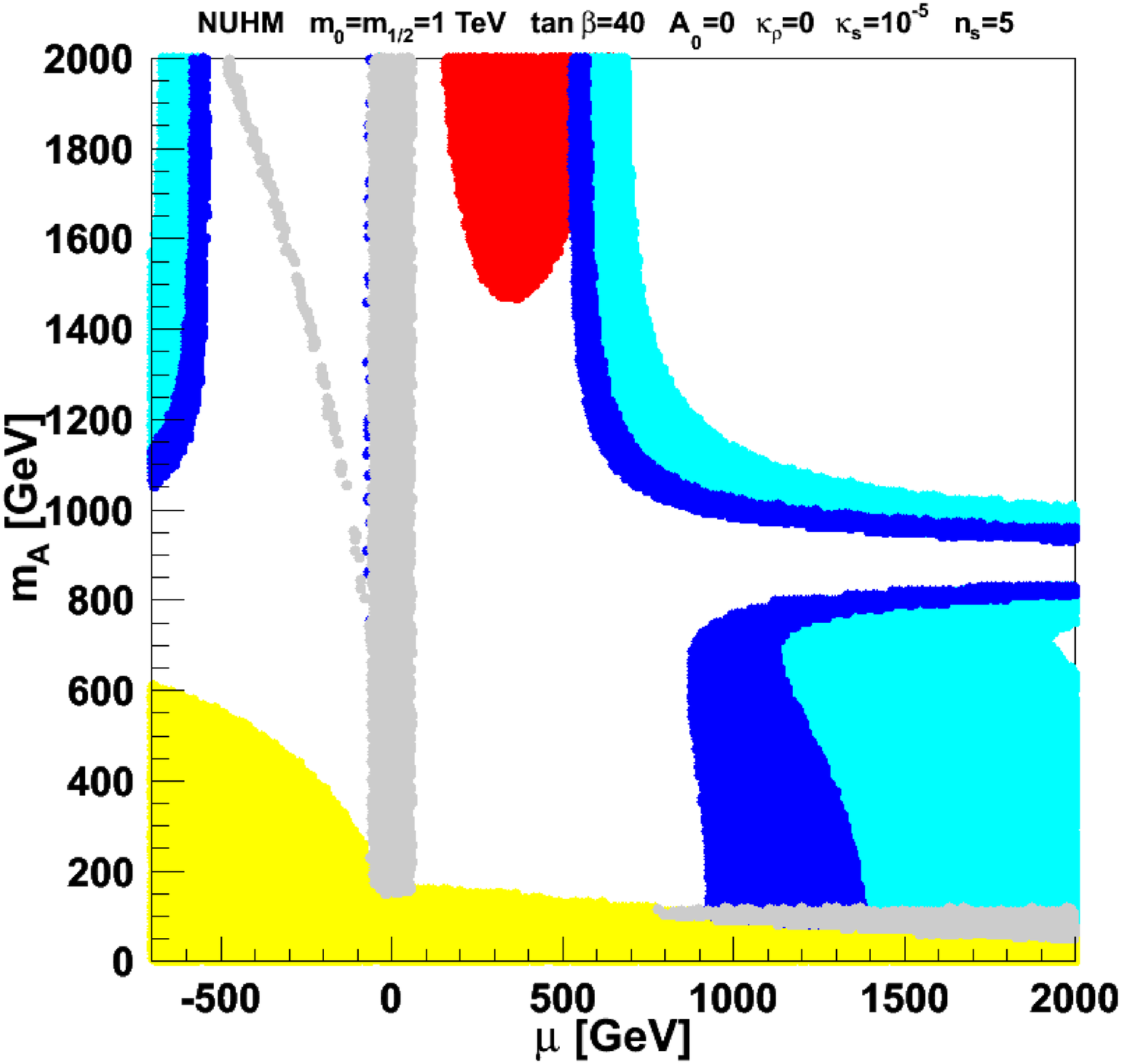}\hspace*{0.2cm}\includegraphics[width=6.45cm]{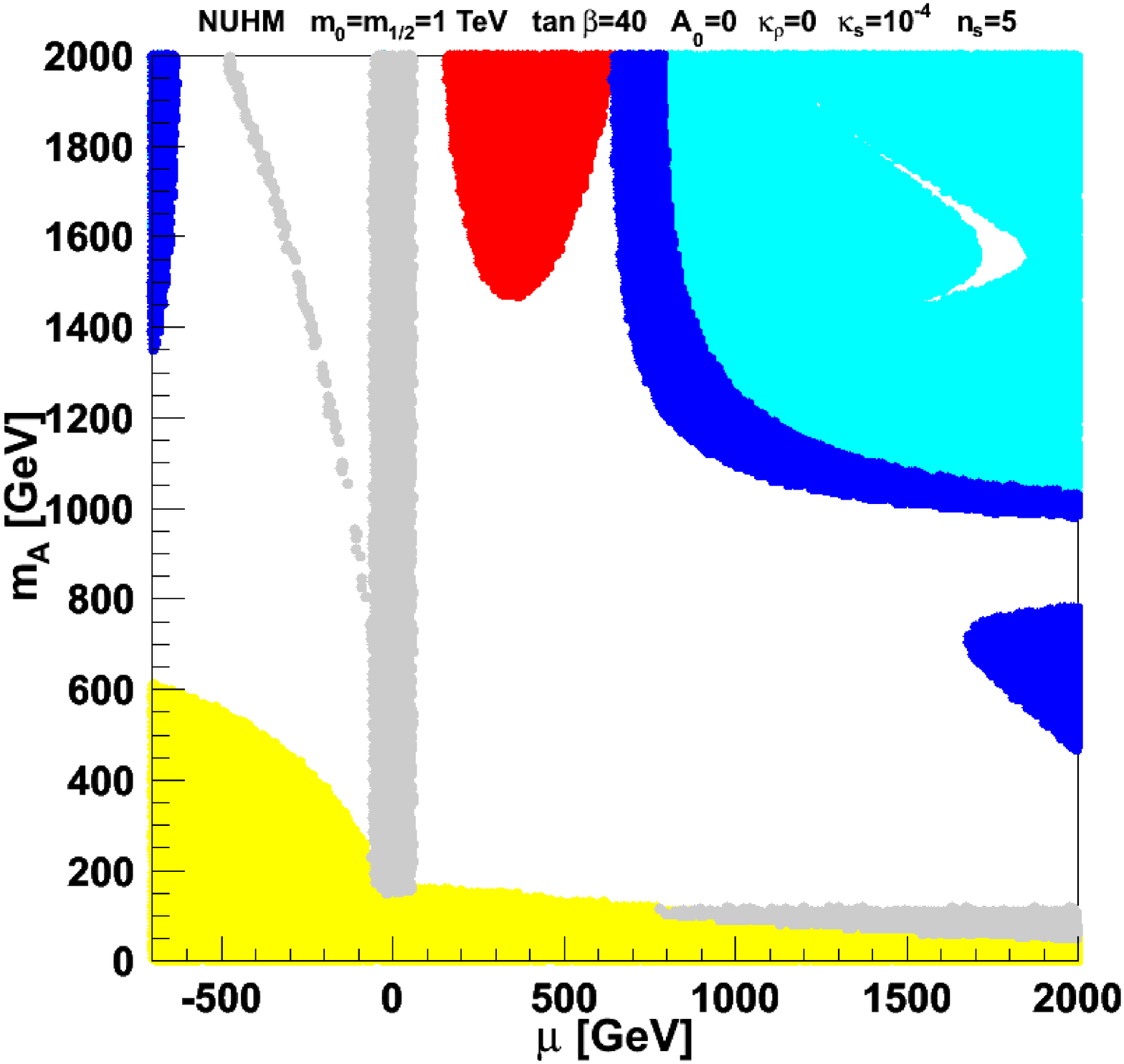}
\end{center}
\caption{Constraints on the NUHM parameter plane ($\mu$,$m_A$), from left to right and top to bottom, in the standard cosmological model, in presence of a tiny energy overdensity with $\kappa_\rho=10^{-4}$ and $n_\rho=6$, in presence of a tiny entropy overdensity with $\kappa_s=10^{-3}$ and $n_s=4$, with $\kappa_s=10^{-2}$ and $n_s=4$, with $\kappa_s=10^{-5}$ and $n_s=5$, and with $\kappa_s=10^{-4}$ and $n_s=5$. The red points are excluded by the isospin asymmetry of $B \to K^* \gamma$, the gray area is excluded by direct collider limits, the yellow zone involves tachyonic particles, and the dark and light blue strips are \underline{favored} by the WMAP constraints and by the older interval (\ref{old}) respectively.\label{fig4}}
\end{figure}%
\begin{figure}[!t]
\begin{center}
\includegraphics[width=7.cm]{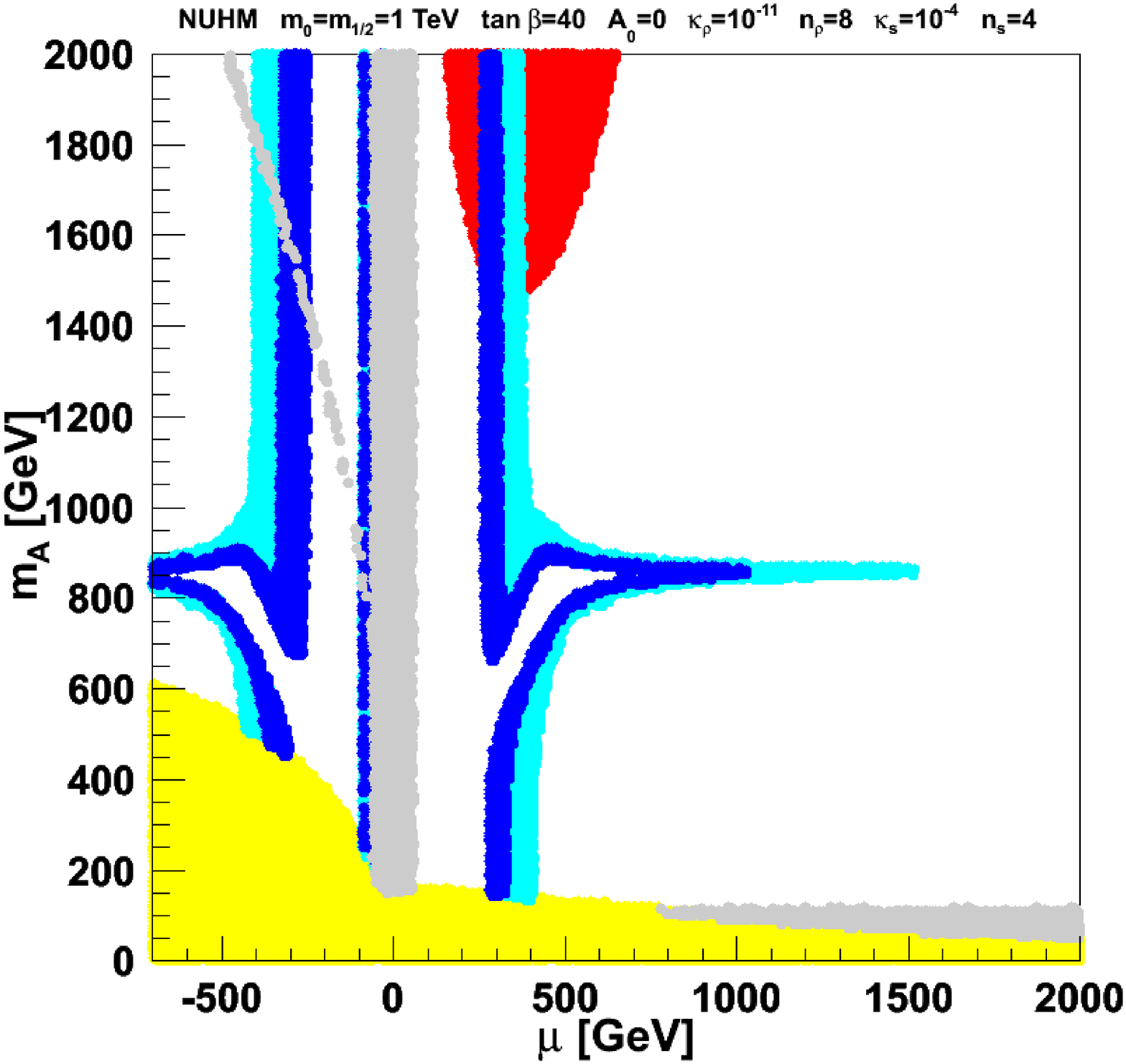}\hspace*{0.2cm}\includegraphics[width=7.cm]{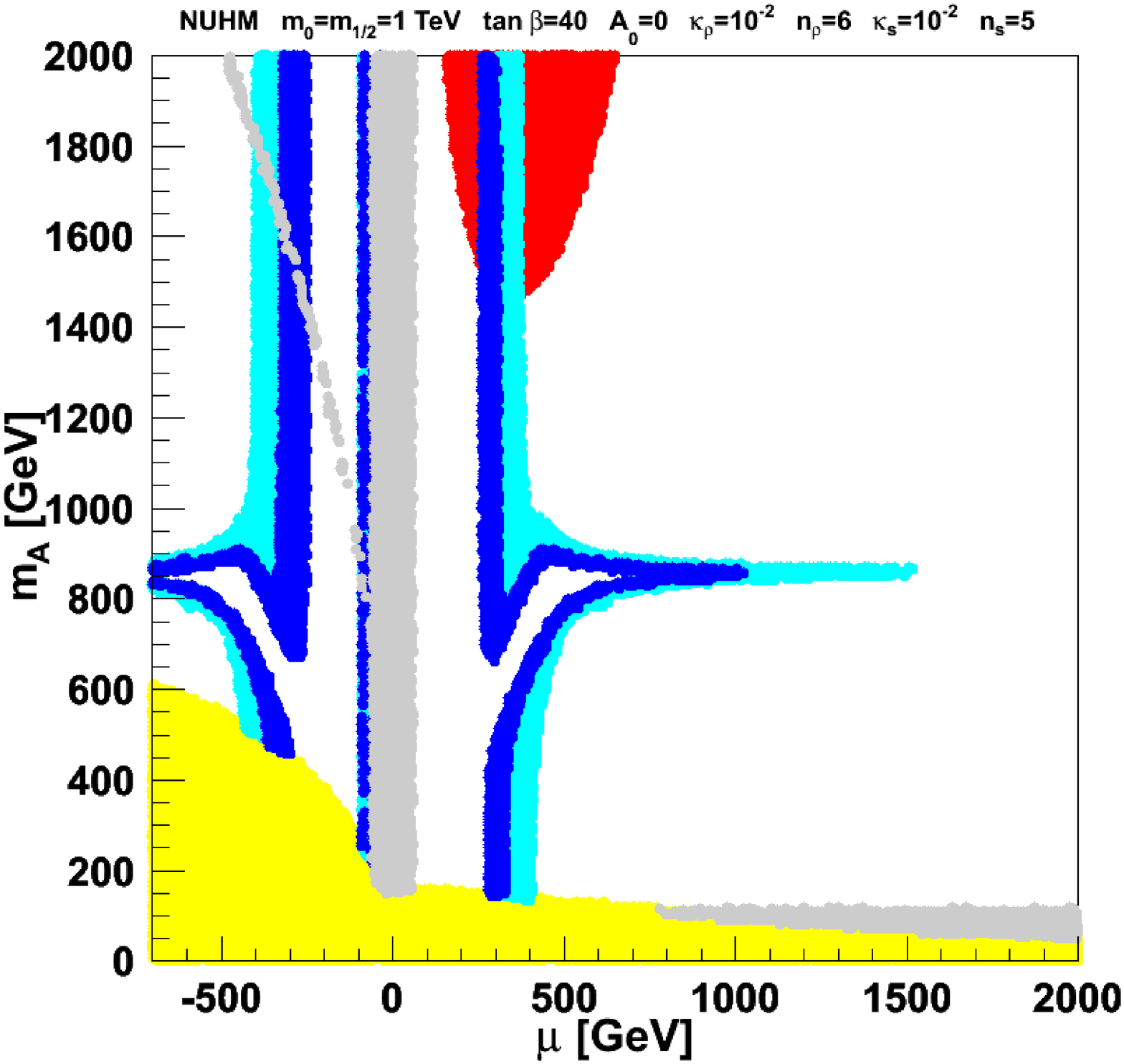}
\end{center}
\caption{Constraints on the NUHM parameter plane ($\mu$,$m_A$), in presence of a tiny energy overdensity with $\kappa_\rho=10^{-11}$ and $n_\rho=8$ together with a tiny entropy overdensity with $\kappa_s=10^{-4}$ and $n_s=4$ on the left, and an energy overdensity with $\kappa_\rho=10^{-2}$ and $n_\rho=6$ as well as an entropy overdensity with $\kappa_s=10^{-2}$ and $n_s=5$ on the right. The colors are as in Fig.~\ref{fig4}.\label{fig5}}
\end{figure}%

We then study the effects of our parametrizations while scanning over the NUHM parameter space. About one million random SUSY points in the NUHM parameter plane ($\mu$,$m_A$) with $m_0=m_{1/2}=1$ TeV, $A_0=0$, $\tan\beta=40$ are generated using SOFTSUSY v2.0.18 \cite{softsusy}, and for each point we compute flavor physics observables, direct limits and the relic density with SuperIso Relic v2.7.

In Fig.~\ref{fig4}, the effects of the cosmological models on the relic density constraints are demonstrated. The first plot is given as a reference for the standard cosmological model, showing the tiny strips corresponding to the regions favored by the relic density constraint. In the second plot, generated in a Universe with an additional energy density with $\kappa_\rho=10^{-4}$ and $n_\rho=6$, the relic density favored strips are reduced, as already shown also in \cite{arbey}, because the calculated relic densities are decreased in comparison to the relic densities computed in the standard scenario. 
The next plots demonstrate the influence of an additional entropy density compatible with BBN constraints. The favored strips are this time enlarged and moved towards the outside of the plot. This effect is due to a decrease in the relic density.

In Fig.~\ref{fig5}, we consider two cosmological scenarios in which both energy and entropy densities are present. The energy and entropy densities have opposite effects and can compensate, and the similarity of the plots reveals the degeneracy between the two cosmological scenarios from the point of view of particle physics. However, using the BBN constraints, the scenario of the right plot can be ruled out.

An important consequence of this example is that if we discover that the particle physics scenario best in agreement with the LHC data (or future colliders) leads to a relic density in disagreement with the cosmological data constraints, useful information on the cosmological scenario may be deduced: first, it would imply that the cosmological standard model does not describe satisfyingly the pre-BBN Universe. Second, combining all cosmological data, and in particular those from BBN, it would be possible to determine physical properties of the Early Universe and constrain Early Universe scenarios. As such, valuable constraints on cosmological models can be obtained from particle colliders.

It is important to point out that all the scenarios previously described -- apart in the second plot of Fig.~\ref{fig5} -- are equivalent from the point of view of the cosmological observations: there is no way to distinguish between them with the current cosmological data. 

\section{Conclusion}

In this paper, we have argued that the calculation of the relic density is very sensitive to the Very Early Universe properties such as the energy or the entropy contents. For this reason, the interpretation of the relic density in the context of particle physics should be done very cautiously, as the favored parameter zones can be completely shifted in any direction by unobservable cosmological phenomena during the pre-BBN era. In \cite{arbey}, we already showed that the addition of energy density can falsify the use of the lower WMAP limit. In this paper we demonstrated that the possible presence of additional entropy also strongly questions the use of the upper WMAP bound. Therefore we can conclude that using the WMAP bounds to constrain SUSY is very model-dependent since the standard model of cosmology remains very uncertain and highly questionable before BBN. On the other hand, we also notice here the importance of the discovery of new physics beyond the particle physics Standard Model: if the LHC data point to a new particle physics model providing a candidate for dark matter, combining relic density calculations with cosmological data will give constraints on the pre-BBN era and give valuable hints on the physics of the Very Early Universe. To conclude, the relic density, even if not predictive for particle physics at the moment, will hopefully soon appear as a new tool to design more precise cosmological models and to analyze the nature of the dark components of the Universe.
%
\subsection*{Acknowledgments}
\noindent A.A. would like to thank K. Jedamzik for his help with the Kawano BBN code.

\end{document}